# Concise Probability Distributions of Eigenvalues of Real-Valued Wishart Matrices

Oliver James and Heung-No Lee

*Abstract*— In this paper, we consider the problem of deriving new eigenvalue distributions of real-valued Wishart matrices that arises in many scientific and engineering applications. The distributions are derived using the tools from the theory of skew symmetric matrices. In particular, we relate the multiple integrals of a determinant, which arises while finding the eigenvalue distributions, in terms of the Pfaffian of skew-symmetric matrices. Pfaffians being the square root of skew symmetric matrices are easy to compute than the conventional distributions that involve Zonal polynomials or beta integrals. We show that the plots of the derived distributions are exactly coinciding with the numerically simulated plots.

## I. Introduction

Eigenvalues of random matrices summarize the macroscopic properties of the engineering and scientific systems in the form of probability distributions. Hence, many researches aim to find the probability distributions of the eigenvalues of the random matrices. In multivariate statistics [1] and in wireless MIMO communications [2], the eigenvalue distributions of random Wishart matrices are widely used for analysis. Let $A_K$ be an $M \times K$ real or complex-valued Gaussian matrix. Then, the $K \times K$ matrix $W = A_K^H A_K$ is called a Wishart matrix, where $A_K^H$ denotes the conjugate transpose of $A_K$.

In many multivariate applications such as financial data analysis, genetic studies, search engines, and climate studies, the study of sample covariance matrices is fundamental and real-valued Wishart matrices are good models of the sample covariance matrices [3]. Hence, the eigenvalues of the real-valued Wishart matrices play a crucial role in multivariate applications. On the other hand, the applications such as MIMO communications require eigenvalues of complex-valued Wishart matrices in order to analyze information theoretic capacity.

In general, a marginal eigenvalue distribution is obtained by integrating the joint eigenvalue distribution over all other eigenvalues except the required one. The joint eigenvalue distribution for the real [4, eq. (58)] and the complex Wishart matrices [4, eq. (95)] have been known for nearly five decades. However, the marginal distributions are known, or can be expressed in tractable form only for some special cases, such as the largest or the smallest eigenvalues.

The first known largest eigenvalue distribution for complex Wishart matrices was derived by Khatri [5] in terms of the product of beta integrals. In [6], Krishnaiah and Chang derived the smallest eigenvalue distribution (for the complex case) in terms of infinite series of Zonal polynomials. An explicit and usable form of Zonal polynomials is available only for small polynomial orders [4, p. 498]. In addition, Zonal polynomials are notoriously difficult to compute. It has been shown in [6] that a series that involves Zonal polynomials converges very slowly. These problems lead to other directions for deriving the eigenvalue distributions during the late 60s. A notable direction was to find a cumulative distribution rather than the probability distribution. Khatri [7] derived a cumulative distribution function for the largest and smallest eigenvalues in terms of a matrix determinant. Ming and Alouini [8] constructed a probability distribution from the cumulative distribution using the idea of derivative of a determinant. However, it has been shown [8, p.1418] that the closed-form probability distribution can be derived only for the special case of a $2 \times 2$ matrix.

Just like the complex-valued case, the extreme eigenvalues of the real-valued Wishart matrices are also derived in terms of Zonal polynomials by Sugiyama and Fukutomi [9]. In [10], Edelman derived the smallest and largest eigenvalue distributions in terms of Tricomi functions. Edelman gave the exact distributions only for a certain matrix sizes leaving behind a recursive expression for computing the distributions [10, p. 45]. The complexity of calculating the recursive functions is questionable for large matrix sizes. In order to alleviate the burden of computing the close-form expression for eigenvalue distributions, limit distributions of the eigenvalues are usually derived [10-14]. These limit distributions show the behavior of the eigenvalues only when the system size tends to infinity and they may not necessarily capture the effects of finite size matrices that occur in practice. Thus, there arises a need to derive novel, tractable eigenvalue distributions for the eigenvalue based analysis of systems.

In [15], Chiani, Win and Zanella attempted to derive new, closed-form expressions for the marginal distributions of eigenvalues for complex Wishart matrices. Their idea is to first express the joint eigenvalue distribution as a product of two determinants. They then derived the marginal distributions by evaluating the multi-dimensional integration of the product of determinants. In [16], Chiani, Win and Zanella again used this same idea to derive the distributions for any arbitrary eigenvalue. Until now, the distributions given by Chiani, Win and Zanella [15, 16] are the best

Oliver is with the Institute of Basic Sciences, Sungkyunkwan University, Suwon, South Korea and Heung-No Lee is with the School of Information and Communications, Gwangju Institute of Science and Technology (GIST), South Korea
(e-mail: j.olivermount@gmailcom; heungno@gist.ac.kr).
This work was supported by the National Research Foundation of Korea grant funded by the Korean government (Do-Yak Research Program, No. 2013-035295).

available closed-form expressions for the complex Wishart matrices. Due to the simplicity and tractability of these distributions, they are conveniently used to evaluate the information theoretic capacity of MIMO systems [16].

It is well-known [1, 10] that the joint eigenvalue distributions for both complex and real-valued Wishart matrices are two different real-valued functions. The way the difference arrived between these two joint eigenvalue distributions from their Wishart matrix distribution is succinctly summarized in Theorem 3.1 [10, p. 33] and Theorem 3.2 [10, p. 36]. Theorem 3.1 discusses the transformations between the volume elements of a Wishart matrix and their eigenvalues. Using these transformations, Theorem 3.2 presents a way to convert the distribution of the Wishart matrices into their joint eigenvalue distribution. Thus, the joint eigenvalue distribution of real-valued Wishart matrices cannot be obtained simply from their complex-valued counterpart. While Chiani, Win and Zanella [16] derived the eigenvalue distributions for the complex-valued case, we derive them, in this paper, for the real-valued case.

In this paper, we aim to derive new eigenvalue distributions for the real-valued Wishart matrices. We derive the distributions borrowing the tools from the theory of skew-symmetric matrices. In particular, we show the eigenvalue distributions in terms of Pfaffian of a skew-symmetric matrix. We observe that unlike the complex-valued case [15, 16], the distributions of eigenvalues for the real-valued case are different for odd and even matrix orders. We show that our distributions can be readily calculated for large matrix orders.

This paper is organized as follows. Section II reviews the joint distribution of the eigenvalues of Wishart matrices and Section III presents the derivation of the extreme eigenvalue distributions. A new eigenvalue framework for the analysis of CS systems is developed in Section IV and Section V concludes the paper.

*Notations:* Capital letters represent matrices while bold face small letter denote vectors. The symbol $|A|$ denotes the matrix determinant if $A$ is a matrix and it denotes the set cardinality if $A$ is a set. The symbol $A^T$ represents the transpose of $A$, $\text{diag}(a_1, a_2, \cdots, a_K)$ denotes a $K \times K$ diagonal matrix with diagonal entries $a_1, a_2, \cdots, a_K$.

## II. Preliminaries

We first review the joint distribution of eigenvalues of the real-valued Wishart matrix [1], which is used to derive the new eigenvalue distributions in the next section.

Let $A$ be an $M \times N$ real-valued Gaussian sensing matrix and $A_K$ be a random collection of $K$ columns of $A$. We assume $K < M < N$. Then, we note that $W := A_K^T A_K$ is a $K \times K$ real-valued Wishart matrix, where $A_K$ is an $M \times K$ real-valued Gaussian matrix. We note that $M > K$ corresponds to a class of full-rank Wishart matrices and we deal with these matrices in this paper. The covariance matrix of each column of $A_K^T$ is given by $\Sigma = \mathrm{E}[\mathbf{a}_i \mathbf{a}_i^T]$. In this paper, we consider an uncorrelated Gaussian matrix, i.e., $\Sigma = \rho I_K$, where $\rho$ is a scalar and the determinant $|\Sigma| = \rho^K$.

Let $\lambda_1 > \lambda_2 > \cdots > \lambda_K$ represent the ordered eigenvalues of the Wishart matrix $W$. We define $\Lambda = \text{diag}(\lambda_1, \lambda_2, \cdots, \lambda_K)$ and thus, $|\Lambda| = \prod_{i=1}^{K} \lambda_i$. The joint distribution of the eigenvalues of a real-valued Wishart matrix is then given by [1, p. 107]

$$f(\lambda) = \frac{\pi^{K^2/2}}{2^{KM/2} \Gamma_K\left(\frac{M}{2}\right) \Gamma_K\left(\frac{K}{2}\right)} \; {}_0F_0\left(-\tfrac{1}{2}\Sigma^{-1}, \Lambda\right) \cdot |\Sigma|^{-\frac{M}{2}} |\Lambda|^{\frac{(M-K-1)}{2}} \prod_{i<j}^{K}(\lambda_i - \lambda_j) , \tag{1}$$

where $\Gamma_p(a)$ is a multivariate gamma function [1] given by

$$\Gamma_p(a) = \pi^{p(p-1)/4} \prod_{i=1}^{p} \Gamma\left(a - \tfrac{1}{2}(i-1)\right).$$

In (1), the term ${}_0F_0\left(-\tfrac{1}{2}\Sigma^{-1}, \Lambda\right)$ represents the hyper-geometric function defined as [1, p. 107]

$${}_0F_0\left(-\tfrac{1}{2}\Sigma^{-1}, \Lambda\right) = \prod_{i=1}^{K} \exp\left(-\tfrac{\lambda_i}{2\rho}\right) \tag{2}$$

Equation (2) is valid only for the case $\Sigma = \rho I_K$. When $\Sigma \neq \rho I_K$, that is, for a correlated case, it is difficult to derive ${}_0F_0\left(-\tfrac{1}{2}\Sigma^{-1}, \Lambda\right)$ [17, p. 445]. However, for some special case of correlated $\Sigma$, (2) can be expressed in terms of an infinite series of zonal polynomials [17]. Since closed-form expression for the hyper-geometric function for correlated case is unavailable, deriving the closed-form marginal distributions of the real-valued correlated Wishart matrices still remains an open problem. On the other hand, for the correlated complex-valued Wishart matrices, the term ${}_0F_0\left(-\tfrac{1}{2}\Sigma^{-1}, \Lambda\right)$ can be evaluated using Harish-Chandra–Itzykson–Zuber integral [17]. In (1), the product term $\prod_{i<j}^{K}(\lambda_i - \lambda_j)$ represents the determinant $|V(\lambda)|$ of a Vandermonde matrix whose $(i,j)$th entry is $v_j(\lambda_i) = \lambda_i^{K-j}$. The matrix $V(\lambda)$ is given below.

$$V(\lambda) = \begin{bmatrix} \lambda_1^{K-1} & \lambda_1^{K-2} & \cdots & 1 \\ \lambda_2^{K-1} & \lambda_2^{K-2} & \cdots & 1 \\ \vdots & \cdots & \cdots & \vdots \\ \lambda_K^{K-1} & \lambda_K^{K-1} & \cdots & 1 \end{bmatrix}. \tag{3}$$

By substituting (2) into (1), the joint density can be written in the form of a matrix determinant as

$$f(\lambda) = \underbrace{\frac{\pi^{K^2/2} \rho^{-KM/2}}{2^{KM/2} \Gamma_K\left(\frac{M}{2}\right) \Gamma_K\left(\frac{K}{2}\right)}}_{c} |V(\lambda)| \prod_{i=1}^{K} \underbrace{\lambda_i^{(M-K-1)/2} e^{\frac{-\lambda_i}{2\rho}}}_{\xi(\lambda_i)}$$

$$= c\,|V(\lambda)| \prod_{i=1}^{K} \xi(\lambda_i) . \tag{4}$$

In (4), we arrange the joint distribution as a product of a constant, a Vandermonde determinant and the product of functions $\xi(\lambda_i)$. We arrange the joint distribution in this form for the ease of deriving the extreme eigenvalue distributions as discussed in the next section. We note that the joint distribution for the complex-valued case [16, p. 1051] is different from the joint distribution (4) for the real-valued case.

III. DERIVATION OF EXTREME EIGENVALUE DISTRIBUTIONS

In this section, we aim to derive the largest and the smallest eigenvalue distributions of a Wishart matrix from the joint distribution. In our approach, we first expand the determinant in (4) by using the generalized determinant expansion. This expansion results in a multi-dimensional integration of a determinant. We perform the multi-dimensional integration using the results from the theory of skew-symmetric matrices.

Let $f(\lambda_1)$ and $f(\lambda_K)$ represent the largest and the smallest eigenvalue distributions of the Wishart matrix, respectively. These distributions can be written in terms of $f(\lambda)$ as the following $K-1$ dimensional ordered integrals:

$$f(\lambda_1) = \int_0^{\lambda_1} \int_0^{\lambda_2} \cdots \int_0^{\lambda_{K-1}} f(\lambda) \, d\lambda_K \cdots d\lambda_3 d\lambda_2 , \quad (5)$$

$$f(\lambda_K) = \int_{\lambda_K}^{\infty} \cdots \int_{\lambda_3}^{\infty} \int_{\lambda_2}^{\infty} f(\lambda) \, d\lambda_1 d\lambda_2 \cdots d\lambda_{K-1} . \quad (6)$$

In order to derive $f(\lambda_1)$ and $f(\lambda_K)$, we substitute $f(\lambda)$ from (4) into (5) and (6) and obtain

$$f(\lambda_1) = c \int_0^{\lambda_1} \int_0^{\lambda_2} \cdots \int_0^{\lambda_{K-1}} |V(\boldsymbol{\lambda})| \prod_{i=1}^{K} \xi(\lambda_i) \, d\lambda_K \cdots d\lambda_3 d\lambda_2 , \quad (7)$$

$$f(\lambda_M) = c \int_{\lambda_K}^{\infty} \cdots \int_{\lambda_3}^{\infty} \int_{\lambda_2}^{\infty} |V(\boldsymbol{\lambda})| \prod_{i=1}^{K} \xi(\lambda_i) \, d\lambda_1 d\lambda_2 \cdots d\lambda_{K-1} . \quad (8)$$

In [9], the $(K-1)$ dimensional integrals in (7) and (8) are evaluated using the theory of Zonal polynomials. In particular, the integrals are transformed into a slowly decaying infinite series of Zonal polynomials. As mentioned in the introduction, an explicit usable form of Zonal polynomials is available only for orders up to 6 [4, p. 498] that makes the distributions highly difficult to compute. In [10], the $(K-1)$ dimensional integrals are evaluated using the theory of differential equations. Specifically, the integrals are shown [10, p. 41] to satisfy the solution of the Tricomi differential equations and thus, the distributions are given in terms of Tricomi functions. However, the exact distributions are given only for the values of $M = K+1, K+2, K+3$. For $M > K+3$, no exact distribution can be found [10, p. 45]. But, in our approach, we use the theory of skew-symmetric matrices to arrive at a solution for the integrals. Unlike [10], in this paper, we are able to give the exact distributions for all values of $K$ and $M$, and our distributions can be computed readily.

The integration in (7) is over the variables $\lambda_2, \cdots, \lambda_K$. However, the integrand, which contains the determinant $|V(\boldsymbol{\lambda})|$ and the product $\prod_{i=1}^{K} \xi(\lambda_i)$, is a function of $\lambda_1, \lambda_2, \cdots, \lambda_K$. In (7), in order to integrate conveniently, we separate the variable $\lambda_1$ both from the determinant as well as from the product. It is easy to separate $\lambda_1$ from the product. In order to separate $\lambda_1$ from the determinant, we use the generalized determinant expansion provided in Lemma 1. A similar procedure can be followed to separate the variable $\lambda_K$ from the integrand in (8), because the integration in (8) is over the variables $\lambda_1, \lambda_2, \cdots, \lambda_{K-1}$.

***Lemma 1 [18]:*** Let $V(\boldsymbol{\lambda})$ be a $K \times K$ matrix with the $(i, j)$th element denoted by $v_j(\lambda_i)$, a function of $\lambda_i$. The determinant $|V(\boldsymbol{\lambda})|$ can be expanded along the *k*th row, $k = 1, 2, \cdots, K$ as

$$|V(\boldsymbol{\lambda})| = \sum_{n=1}^{K} (-1)^{n+k} v_n(\lambda_k) |V_{nk}(\boldsymbol{\lambda})|, \quad (9)$$

where $V_{nk}(\boldsymbol{\lambda})$ is an $(K-1) \times (K-1)$ sub-matrix obtained by deleting the *k*th row and *n*th column of $V(\boldsymbol{\lambda})$. The sub-matrices $V_{nk}(\boldsymbol{\lambda})$, $n = 1, 2, \cdots, K$, are independent of $\lambda_k$. ∎

In order to understand Lemma 1, we consider an example with $K = 3$ and $v_j(\lambda_i) = \lambda_i^{K-j}$. The matrix $V(\boldsymbol{\lambda})$ and its determinant expansion along the first row $(k = 1)$ are given respectively as

$$V(\boldsymbol{\lambda}) = \begin{bmatrix} \lambda_1^2 & \lambda_1 & 1 \\ \lambda_2^2 & \lambda_2 & 1 \\ \lambda_3^2 & \lambda_3 & 1 \end{bmatrix},$$

And

$$|V(\boldsymbol{\lambda})| = \lambda_1^2 \begin{vmatrix} \lambda_2 & 1 \\ \lambda_3 & 1 \end{vmatrix} - \lambda_1 \begin{vmatrix} \lambda_2^2 & 1 \\ \lambda_3^2 & 1 \end{vmatrix} + 1 \begin{vmatrix} \lambda_2^2 & \lambda_2 \\ \lambda_3^2 & \lambda_3 \end{vmatrix}$$
$$= \sum_{n=1}^{3} (-1)^{n+1} \lambda_1^{3-n} |V_{n1}(\boldsymbol{\lambda})| \quad (10)$$

In (10), the sub-matrices $V_{n1}(\boldsymbol{\lambda})$, $n = 1, 2, 3$ are functions of the variables $\lambda_2$ and $\lambda_3$ only and independent of $\lambda_1$. Thus, we pulled the variable $\lambda_1$ out from the determinant expansion. ∎

Using Lemma 1, we can pull $\lambda_1$ or $\lambda_K$ out from the determinant expansion. Expanding $|V(\boldsymbol{\lambda})|$ in (7) along the first row for the largest eigenvalue $\lambda_1$, and by moving $\xi(\lambda_1)$ outside of the product, we obtain

$$f(\lambda_1) = c \int_0^{\lambda_1} \int_0^{\lambda_2} \cdots \int_0^{\lambda_{K-1}} \sum_{n=1}^{K} (-1)^{n+1} \lambda_1^{K-n} \xi(\lambda_1)$$
$$\cdot |V_{n1}(\boldsymbol{\lambda})| \prod_{i=2}^{K} \xi(\lambda_i) \, d\lambda_K \cdots d\lambda_3 \, d\lambda_2 . \quad (11)$$

In (11), for each $n$, we can combine the product $|V_{n1}(\boldsymbol{\lambda})|\prod_{i=2}^{K}\xi(\lambda_i)$ into a single term by using the scaling property of the determinant, i.e.,

$$|B_n(\boldsymbol{\lambda})| = |V_{n1}(\boldsymbol{\lambda})|\prod_{i=2}^{K}\xi(\lambda_i) . \quad (12)$$

The scaling property dictates that for two matrices, $B$ and $V$, and a constant $\xi$, if $|B| = \xi|V|$, then the matrix $B$ can be obtained by multiplying a row or a column of $V$ by the same constant $\xi$. In the R.H.S of (12), we have a determinant which is multiplied by a product. Using the scaling property, the $(K-1)\times(K-1)$ matrix $B_n(\boldsymbol{\lambda})$ can then be obtained by multiplying $k$th row of the matrix $V_{n1}(\boldsymbol{\lambda})$ with the $k$th term of the product. We illustrate (12) with the example in (10). The product $|V_{n1}(\boldsymbol{\lambda})|\prod_{i=2}^{K}\xi(\lambda_i)$ for $n=1$ is given by

$$|B_1(\boldsymbol{\lambda})| = |V_{11}(\boldsymbol{\lambda})|\prod_{i=2}^{K}\xi(\lambda_i) . \quad (13)$$

From (10), $|V_{11}(\boldsymbol{\lambda})| = \begin{vmatrix} \lambda_2 & 1 \\ \lambda_3 & 1 \end{vmatrix}$ and thus

$$\begin{aligned} |B_1(\boldsymbol{\lambda})| &= \begin{vmatrix} \lambda_2 & 1 \\ \lambda_3 & 1 \end{vmatrix}\xi(\lambda_2)\xi(\lambda_3) \\ &= \begin{vmatrix} \lambda_2\xi(\lambda_2) & \xi(\lambda_2) \\ \lambda_3\xi(\lambda_3) & \xi(\lambda_3) \end{vmatrix}, \end{aligned} \quad (14)$$

where in the second step of (14) the terms $\xi(\lambda_2)$ and $\xi(\lambda_3)$ multiply the first and second rows of $V_{11}(\boldsymbol{\lambda})$, respectively. ∎

Using (12), we rewrite (11) after interchanging the summation and integration as

$$\begin{aligned} f(\lambda_1) &= c\sum_{n=1}^{K}(-1)^{n+1}\lambda_1^{K-n}\xi(\lambda_1) \\ &\quad \cdot \int_0^{\lambda_1}\int_0^{\lambda_2}\cdots\int_0^{\lambda_{K-1}}|B_n(\boldsymbol{\lambda})|\,d\lambda_K\cdots d\lambda_3\,d\lambda_2 . \end{aligned} \quad (15)$$

We will continue to work with (15), the distribution of the largest eigenvalue. For the smallest eigenvalue $\lambda_K$, we follow the similar procedure later on. In (15), for each value of $n$, we need to find a $K-1$ dimensional integration of the determinant of an $(K-1)\times(K-1)$ matrix $B_n(\boldsymbol{\lambda})$. This $K-1$ dimensional integration for an $n$ can be evaluated using a result from the theory of skew-symmetric matrices provided in the following lemma. A square matrix $A$ is skew-symmetric if $A = -A^T$. This implies that the $(i,j)$th element of $A$ satisfies $a_{i,j} = -a_{j,i}$, and the diagonal elements are zero, i.e., $a_{i,i} = 0$. A $K \times K$ matrix is also called a matrix of order $K$.

**Lemma 2 [19, N. G. De Bruijn]:** Let $\theta_i(\lambda_j)$ denote the $(i,j)$th entry of the $(K-1)\times(K-1)$ matrix $B_n(\boldsymbol{\lambda})$. Then,

$$\int_0^{\lambda_1}\int_0^{\lambda_2}\cdots\int_0^{\lambda_{K-1}}|B_n(\boldsymbol{\lambda})|\,d\lambda_K\cdots d\lambda_3 d\lambda_2 = \text{PF}(\mathrm{B}_n), \quad (16)$$

where $\text{PF}(\mathrm{B}_n)$ is the Pfaffian of a $(K-1)\times(K-1)$ skew-symmetric matrix $\mathrm{B}_n$. The entries of $\mathrm{B}_n$ is calculated using the diagonal entries of the matrix $B_n(\boldsymbol{\lambda})$. This calculation is different for odd and even $K$.

<u>Odd $K$</u>: In this case, the matrix $\mathrm{B}_n$ is of order $K-1$. The $(i,j)$th element of $\mathrm{B}_n$ is given by

$$b_{i,j} = \int_0^{\lambda_1}\int_0^{\lambda_1}\theta_i(\lambda_i)\theta_j(\lambda_j)\,\text{sgn}(\lambda_j - \lambda_i)\,d\lambda_i d\lambda_j , \quad (17)$$

for $i,j = 1,2,\cdots,K-1$.

<u>Even $K$</u>: In this case, the matrix $\mathrm{B}_n$ is of order $K+1$. In addition to (17), $\mathrm{B}_n$ has two more rows at the bottom and two more columns at the right whose non-zero values are calculated as

$$\begin{aligned} b_{i,K} &= 1; & i &= 1,2,\cdots,K-1 \\ b_{i,K+1} &= \int_0^{\lambda_1}\theta_i(\lambda)\,d\lambda; & i &= 1,2,\cdots,K \end{aligned} \quad (18)$$

∎

From Lemma 2, we note that the $K-1$ dimensional integration of $|B_n(\boldsymbol{\lambda})|$ is reduced to the calculation of Pfaffian of the skew-symmetric matrix $\mathrm{B}_n$, whose entries can be obtained by using (17) and (18). The $\text{PF}(\mathrm{B}_n)$ can be obtained using the relation $\text{PF}(\mathrm{B}_n) = \sqrt{\det \mathrm{B}_n}$ as $\det \mathrm{B}_n \geq 0$ for skew-symmetric matrices. Thus, applying Lemma 2 to (15), we can find the distribution of the largest eigenvalue in terms of $\text{PF}(\mathrm{B}_n)$. Similarly, an expression for the distribution of smallest eigenvalue $\lambda_K$ can be obtained by expanding the determinant in (8) along the last row and then defining a determinant $|D_n(\boldsymbol{\lambda})| = |V_{nK}(\boldsymbol{\lambda})|\prod_{i=1}^{K-1}\xi(\lambda_i)$ as

$$\begin{aligned} f(\lambda_K) &= c\sum_{n=1}^{K}(-1)^{n+K}\lambda_1^{K-n}\xi(\lambda_K) \\ &\quad \cdot \int_{\lambda_K}^{\infty}\cdots\int_{\lambda_3}^{\infty}\int_{\lambda_2}^{\infty}|D_n(\boldsymbol{\lambda})|\,d\lambda_1 d\lambda_2\cdots d\lambda_{K-1} . \end{aligned} \quad (19)$$

To find the $K-1$ dimensional integration in (19), Lemma 2 can still be used except that the interval of the integration in in Lemma 2 ranges from $\lambda_K$ to $\infty$. In the subsequent discussions, we find the exact distribution for the largest and smallest eigenvalues.

### A. Largest eigenvalue distribution

Using Lemma 2 in (15), the distribution of the largest eigenvalue $f(\lambda_1)$ is given by

$$f(\lambda_1) = c\sum_{n=1}^{K}(-1)^{n+1}\lambda_1^{K-n}\xi(\lambda_1)\operatorname{PF}(B_n)$$
$$= c\sum_{n=1}^{K}(-1)^{n+1}\lambda_1^{K-n+(M-K-1)/2}e^{-\frac{\lambda_1}{2\rho}}\operatorname{PF}(B_n). \quad (20)$$

In (20), in order to arrive at the second step, we substitute the expression $\xi(\lambda_1) = \lambda_1^{(M-K-1)/2}e^{-\frac{\lambda_1}{2\rho}}$. To evaluate $\operatorname{PF}(B_n)$ for each value of $n$, we need to first construct $B_n$ by calculating its entries $b_{i,j}$ s given in (17) and (18). Evaluation of $b_{i,j}$ needs $\theta_i(\lambda_i)$, the diagonal element of $B_n(\lambda)$. The diagonal element is given by $\theta_i(\lambda_i) = \lambda_i^{r_{n,i}}\xi(\lambda_i) = \lambda_i^{r_{n,i}}\lambda_i^{(M-K-1)/2}e^{-\frac{\lambda_i}{2\rho}}$, where $r_{n,i}$ is defined as follows

$$r_{n,i} := \begin{cases} K-i, & \text{if } i < n \\ K-1-i, & \text{if } i \geq n \end{cases}. \quad (21)$$

By substituting the value of $\theta_i(\lambda_i)$ in (17), we obtain

$$b_{i,j} = \int_0^{\lambda_1}\int_0^{\lambda_1}\lambda_i^{r_{n,i}}\lambda_i^{(M-K-1)/2}e^{-\frac{\lambda_i}{2\rho}} \\ \cdot \lambda_j^{r_{n,j}}\lambda_j^{(M-K-1)/2}e^{-\frac{\lambda_j}{2\rho}}\operatorname{sgn}(\lambda_j - \lambda_i)d\lambda_i d\lambda_j. \quad (22)$$

We now use the definition of the sgn(.) function and break the inner integration in (22) into two parts to obtain

$$b_{i,j} = \int_0^{\lambda_1}\lambda_j^{r_{n,j}}\lambda_j^{(M-K-1)/2}e^{-\frac{\lambda_j}{2\rho}} \\ \cdot\left(\int_0^{\lambda_j}\lambda_i^{r_{n,i}}\lambda_i^{(M-K-1)/2}e^{-\frac{\lambda_i}{2\rho}}d\lambda_i \right. \\ \left. -\int_{\lambda_j}^{\lambda_1}\lambda_i^{r_{n,i}}\lambda_i^{(M-K-1)/2}e^{-\frac{\lambda_i}{2\rho}}d\lambda_i\right)d\lambda_j \quad (23) \\ = \frac{1}{\left(\frac{1}{2\rho}\right)^{r_{n,i}+\frac{M-K+1}{2}}}\int_0^{\lambda_1}\lambda_j^{r_{n,j}}\lambda_j^{(M-K-1)/2}e^{-\frac{\lambda_j}{2\rho}} \\ \cdot\left(2\gamma\left(r_{n,i}+\frac{M-K+1}{2},\frac{\lambda_j}{2\rho}\right)-\gamma\left(r_{n,i}+\frac{M-K+1}{2},\frac{\lambda_1}{2\rho}\right)\right)d\lambda_j,$$

where $\gamma(v,\mu a) = \mu^v\int_0^a x^{v-1}e^{-\mu x}dx$ is the lower incomplete gamma integral. Similarly, substituting $\theta_i(\lambda_i)$ in (18), we get

$$b_{i,K+1} = \int_0^{\lambda_1}\lambda^{r_{n,i}}\lambda^{(M-K-1)/2}e^{-\frac{\lambda}{2\rho}}d\lambda \\ = (2\rho)^{r_{n,i}+\frac{M-K+1}{2}}\gamma\left(r_{n,i}+\frac{M-K+1}{2},\frac{\lambda_1}{2\rho}\right) \quad (24)$$

By evaluating the integrals in (23) and (24), we obtain the entries of the matrix $B_n$. The value of $\operatorname{PF}(B_n)$ can then be calculated using the relation $\operatorname{PF}(B_n) = \sqrt{\det B_n}$.

### B. Smallest eigenvalue distribution

Similarly, in order to derive $f(\lambda_K)$, we use Lemma 2 in (19) and obtain

$$f(\lambda_K) = c\sum_{n=1}^{K}(-1)^{n+K}\lambda_K^{K-n}\xi(\lambda_K)\operatorname{PF}(D_n) \\ = c\sum_{n=1}^{K}(-1)^{n+K}\lambda_K^{K-n+(M-K-1)/2}e^{-\frac{\lambda_K}{2\rho}}\operatorname{PF}(D_n) \quad (25)$$

The entries of $D_n$ can be calculated similar to (22) and (24) expect that the interval of the integration is now from $\lambda_K$ to $\infty$ as given below.

$$d_{i,j} = \frac{1}{\left(\frac{1}{2\rho}\right)^{r_{n,i}+\frac{M-K+1}{2}}}\int_{\lambda_K}^{\infty}\lambda_j^{r_{n,j}}\lambda_j^{(M-K-1)/2}e^{-\frac{\lambda_j}{2\rho}} \\ \cdot\left(\Gamma\left(r_{n,i}+\frac{M-K+1}{2},\frac{\lambda_K}{2\rho}\right)-2\Gamma\left(r_{n,i}+\frac{M-K+1}{2},\frac{\lambda_j}{2\rho}\right)\right)d\lambda_j \quad (26)$$

where $\Gamma(v,\mu a) = \mu^v\int_a^{\infty}x^{v-1}e^{-\mu x}dx$ is the upper incomplete gamma integral. Similarly, the value of $d_{i,K+1}$ is

$$d_{i,K+1} = \int_{\lambda_K}^{\infty}\lambda^{r_{n,i}}\lambda^{(M-K-1)/2}e^{-\frac{\lambda}{2\rho}}d\lambda \\ = (2\rho)^{r_{n,i}+\frac{M-K+1}{2}}\Gamma\left(r_{n,i}+\frac{M-K+1}{2},\frac{\lambda_K}{2\rho}\right) \quad (27)$$

In summary, we have pursued the following steps in order to obtain the largest eigenvalue distribution in (20) from (11):

1. Find $B_n(\lambda)$ in (12) using the scaling property of determinants.
2. Obtain $B_n$ using the diagonal entries of $B_n(\lambda)$ via (17) and (18).
3. Obtain the result of the $K-1$ dimensional integration of $|B_n(\lambda)|$ as a Pfaffian of $B_n$ via $\operatorname{PF}(B_n) = \sqrt{\det B_n}$.

We followed the similar steps in order to obtain the smallest eigenvalue distribution in (25) from (19).

## IV. SIMULATION RESULTS AND DISCUSSIONS

In this section, we aim to verify the exactness of the probability distributions in (20) and (25) via numerically finding their empirical distributions (explained below). The plots of the empirical as well as its analytical distribution in (20) for the largest eigenvalue distribution are plotted in Fig. 1 for $K=51$ and for various values of $M$ with $\rho=\frac{1}{M}$. Similar plots for the smallest eigenvalue distribution are shown in Fig. 2. In Figs. 1 and 2 the empirical plots are shown in black color dashed curves and the plots of the analytical curves are shown in solid curves. The empirical distributions are calculated as follows:

1. We generate an i.i.d. Gaussian matrix $A$ of size $M \times 5000$ for $M = 300, 500,$ and $700$.
2. We randomly draw a sub-matrix $A_K$ of size $M \times K$ from $A$ and find the largest and smallest eigenvalues of the matrix $A_K^T A_K$ using the MATLAB command *eig*. We call these values as the empirical values.
3. We repeat step 2 for 10000 times and obtain 10000 empirical values from which we compute the empirical probability distributions.

It is evident from Figs. 1 and 2 that the empirical distributions exactly coincide with the analytical distributions give in Eqs. (20) and (25).

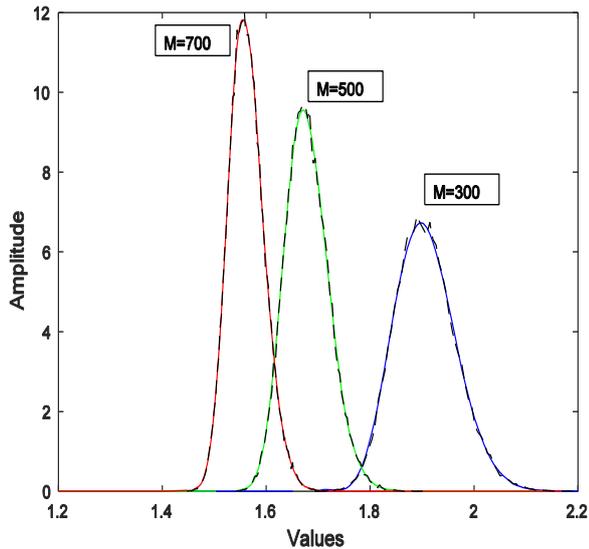

Fig. 1. Largest eigenvalue distribution of Wishart matrix for various *M*.

The large deviation result [12] says that as $M$ increases, 1) the largest and the smallest eigenvalues are found near to 1, and 2) the shapes of the eigenvalues get sharper. This is confirmed from the plots in Figs. 1 and 2 in which as *M* increase the eigenvalues move closer to 1 while their shapes get sharper.

The eigenvalue distributions in [10] are the most widely used in various applications. However, our probability distribution expressions are readily computable compared to the results in [10]. For example, the distributions in [10, p. 53] are plotted for $K = 3$ and $M = 27$. However, in this paper, we plot them for $K = 51$ and $M = 700$. As per our knowledge, we have not seen the eigenvalue distributions computed and plotted for the range of values we have shown in these figures. For values of $K$ and $M$ much larger than that of those shown in the figures, the computation of the distributions is still difficult due to the numerical evaluation steps, i.e., (23)-(24) and (26)-(27). Our future research goal is to overcome this issue in the computation.

## V. CONCLUSIONS

In this paper, we have derived new closed-form eigenvalue distributions for real-valued Wishart matrices. The distributions have been derived by using the results from the theory of skew-symmetric matrices. In particular, we have shown that unlike the complex case the distributions of the eigenvalues for the real-valued matrices are different for odd and even size of the Wishart matrix. We have demonstrated the correctness of the derived distributions by comparting with their empirical distributions. Unlike the distributions in the literature which are in terms of complex Zonal polynomials and beta integrals, our distributions are more concise and can be readily computed.

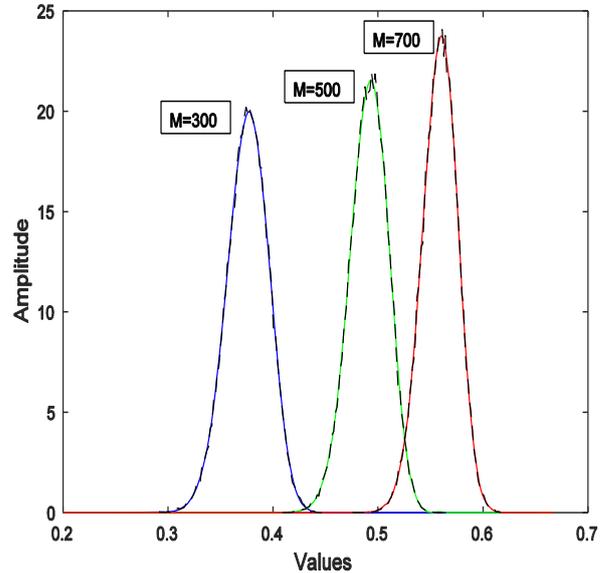

Fig. 2. Smallest eigenvalue distribution of Wishart matrix for various *M*.